\documentclass[11pt,a4paper]{article}
\usepackage[left=15mm, top=10mm, right=15mm, bottom=20mm, nohead=true, nofoot=true]{geometry}

\usepackage[utf8]{inputenc}
\usepackage{authblk}
\usepackage{indentfirst}
\usepackage{misccorr}
\usepackage{graphicx}
\usepackage{amsmath}
\usepackage{amssymb}
\usepackage{multicol}
\usepackage{bm}
\usepackage{longtable}
\usepackage{pdflscape}
\usepackage{epstopdf}
\usepackage{multirow}
\usepackage{adjustbox}
\usepackage{amsfonts}
\usepackage{ifthen}
\usepackage{fancyhdr}
\usepackage{setspace}
\usepackage{xcolor}
\usepackage[round,authoryear,comma,sort&compress,numbers]{natbib}
\usepackage[toc,title,page]{appendix}
\usepackage[hidelinks]{hyperref}
\usepackage{url}

\hypersetup{
    colorlinks=true,
    linkcolor=green,
    citecolor=blue,
    filecolor=magenta,
    urlcolor=cyan,
    pdftitle={Sharelatex Example},
    pdfpagemode=FullScreen,
}

\fancypagestyle{plain}{\chead{\texttt{\textcolor{brown}{Communications of BAO, Vol.\,72, Issue\,2, 2025, pp.\,334-340}}}}
\title{\textbf{Website with interactive visualization of multivariate astronomical time series}}
\author[1]{M. Volkov$\star$}
\author[2,3]{M. Demianenko$\star$\thanks{$\star$ these authors made equal contribution; demianenko@mpia.de, corresponding author;\\ Fellow of the International Max Planck Research School for Astronomy and Cosmic Physics at the University of Heidelberg (IMPRS-HD).}}
\author[4]{D. Matveev}
\author[5,4]{K. A. Grishin}
\author[6,4]{I. V. Chilingarian}

\affil[1]{\scriptsize Yandex LLC, 16 Lev Tolstoy Street, 119021, Moscow, Russia}
\affil[2]{\scriptsize Max-Planck-Institut für Astronomie, Königstuhl 17, 69117 Heidelberg, Germany}
\affil[3]{\scriptsize Department for Physics and Astronomy, Heidelberg University, Im Neuenheimer Feld 226, 69120 Heidelberg, Germany}
\affil[4]{\scriptsize Sternberg Astronomical Institute, Moscow M.V. Lomonosov State University, Universitetskij pr., 13,  Moscow, 119234, Russia}
\affil[5]{\scriptsize Universit\'e Paris Cit\'e, CNRS(/IN2P3), Astroparticule et Cosmologie, F-75013 Paris, France}
\affil[6]{\scriptsize Center for Astrophysics --- Harvard and Smithsonian, 60 Garden Street MS09, Cambridge, MA 02138, USA}

\begin{document}
\pagestyle{empty}
\newpage
\pagestyle{fancy}
\label{firstpage}
\date{}
\maketitle

\begin{abstract}
Light curves represent astronomical time series of flux measured across one or more photometric bands. With the rapid growth of large-scale sky surveys, time-domain astronomy has become an essential area of modern astrophysical research. Interactive visualization of extensive light-curve datasets plays a key role in exploring transient phenomena and in planning large follow-up campaigns. In this work, we introduce two web-based platforms designed for interactive light-curve visualization: {\sc Fulu}, for transient event studies, and VALC, for investigations of low-mass active galactic nuclei (AGNs). These tools provide a user-friendly interface for examining, comparing, and interpreting vast collections of astronomical light curves, supporting scientific discovery.
\end{abstract}

\section{Motivation}  
Interactive visualization of multivariate time series in astronomy, specifically light curves of astrophysical sources, is an important and often missing component of wide-field sky surveys and also of smaller, focused astronomical databases. Keeping this in mind, we developed the website\footnote{\url{https://lc-dev.voxastro.org}} for light curve visualization and exploration.  Our platform allows users to explore and visualize the results of {\sc Fulu} neural network approximation and variability analyzer for light curves (VALC) algorithms applied to real observational samples, offering object-by-object insights into time-domain behavior. In this paper, we present the website’s architecture, core functionality, and its scientific applications, demonstrating how it facilitates efficient analysis and interpretation of complex light-curve data.
\subsection{FULU}
{\sc Fulu}~\citep{2023A&A...677A..16D} is a Python package that provides ready-to-use neural network and Gaussian processes~\citep[GP;][]{NIPS1995_7cce53cf} approximation methods for multiband light curves with uneven time steps, accessing approximation both through time and passband. Implemented methods (Bayesian neural network~\citep{blundell2015weight}, normalizing flows~\citep{NF1, NF2}, multilayer perceptions~\citep{1986Natur.323..533R} with one and two hidden layers, GP) predict the observations and their errors in several photometric bands. The models require at least 10 points within one light curve for fair performance.
The tests on the Zwicky Transient Facility~\citep[ZTF;][]{2019PASP..131a8003M} Bright Transient Survey~\citep[BTS;][]{2020ApJ...895...32F,2020ApJ...904...35P} light curves based on {\sc Fulu} showed that neural network models are a reasonable alternative to GP and also take less computational time than the widely-used GP with different covariance matrices~\citep{2023JPhCS2438a2128D}. Both PLAsTiCC Legacy Survey of Space and Time~\citep[LSST;][]{2023PASP..135j5002H} simulated~\citep{2023ApJS..267...25H} and real ZTF BTS datasets were validated by regression metrics, quality metrics of
observations predicted by the models, as well as the quality of subsequent classification and peak estimation~\citep{2024ASPC..535..131S,2023A&A...677A..16D,2023yCat..36770016D}. 

We present the {\sc Fulu} webpage\footnote{\url{https://lc-dev.voxastro.org/fulu.html}} with the BTS ZTF sample of neural network and GP approximated transient light curves, with estimated peak time and magnitude. 
\subsection{VALC}
Asteroid Terrestrial-impact Last Alert System \citep[ATLAS;][]{2018PASP..130f4505T} Forced Photometry service~\citep[AFPS;][]{2021TNSAN...7....1S,2020PASP..132h5002S} provides PSF-photometry at different images adopting the algorithm {\sc tphot}~\citep{2011PASP..123...58T,2013PASP..125..456S}. 
The ZTF Forced Photometry service~\citep[ZFPS][]{2023arXiv230516279M} produces \textit{statistically optimal} PSF-photometry at different imaging~\citep[ZOGY;][]{2016ApJ...830...27Z} and the adoption of it for aperture photometry. Such light curves enable us to investigate the variability amplitudes of active galactic nuclei (AGN), eliminating the host galaxy contribution. However, even in the case of the ZOGY, we still observe artificial variability contributions in ZFPS, the exact source of which has not yet been identified. \citet{2024ASPC..535..283D,2022aems.conf..359D} developed a post-processing algorithm for the artificial variability elimination in ZFPS light curves using non-variable stars, then applied it to $\sim$1900 low-mass AGN candidates up to $2\times10^{6}$ $M_{\odot}$ selected by broad optical emission lines using SDSS DR7 spectra~\citep{2017ApJS..228...14C,2018ApJ...863....1C}.

We developed a VALC webpage\footnote{\url{https://lc-dev.voxastro.org/valc.html}} with ZTF and ATLAS Forced Photometry light curves and main parameters of these objects. 

\section{Architecture and functionality}
\begin{figure}
\centering
\includegraphics[width=.7\textwidth]{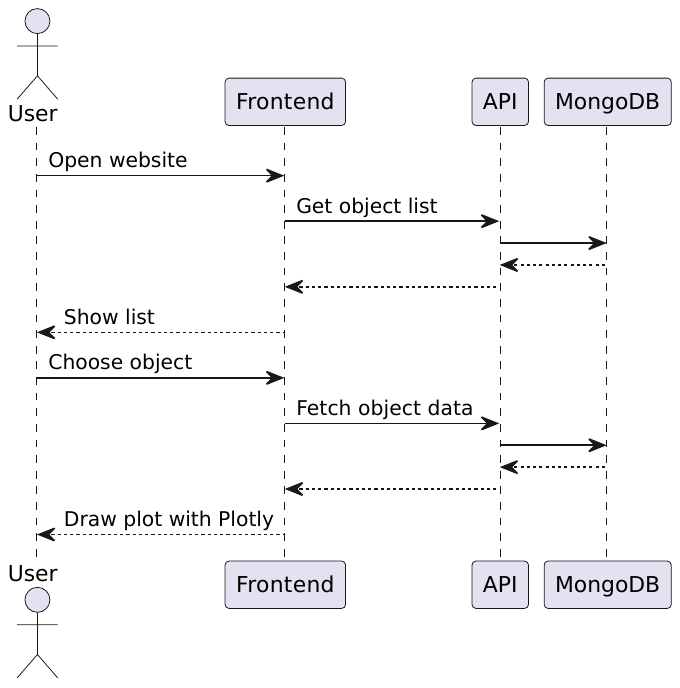}{(a)}
\includegraphics[width=0.5\textwidth]{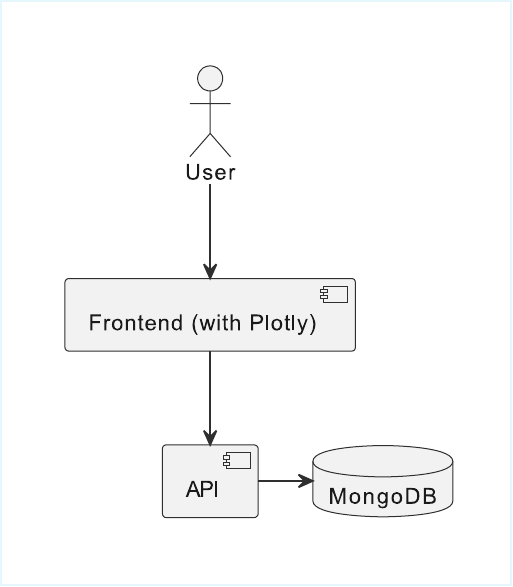}{(b)}
\caption{(a) Sequence diagram illustrating the data flow between the user, frontend, API, and MongoDB when displaying and plotting object data using Plotly\label{fig: seq}. (b) Component diagram illustrating simple architecture of the system.\label{fig: comp}}
\end{figure}

The website features visualizations that utilize {\sc Plotly}~\citep{plotly} to display interactive graphics. On the server side, the system is implemented as a Python application using the {\sc FastAPI}~\citep{fastapi} framework. Data are stored as JSON~\citep{crockford2006json} objects in MongoDB~\citep{mongodb}. All components of the system, including the frontend, backend, and database, are encapsulated within Docker~\citep{merkel2014docker} containers. Figure~\ref{fig: seq}(a) shows process interactions between the user, frontend, API, and MongoDB arranged in a time sequence. The system architecture is shown in Figure~\ref{fig: comp}(b). 

Such an architecture was chosen for its simplicity and ease of maintenance. Docker containers enable full system deployment on any server with a few commands. FastAPI, with its strong typing support, provides strict control over data models while maintaining high development speed in Python. The client is separated from the backend, allowing for future migration to another frontend technology without affecting the backend - a step we plan to take as the current frontend becomes harder to maintain. Since the system does not involve complex data filtering logic, a document-oriented database such as MongoDB was found to be a suitable solution.

Data ingestion to the website occurs in two stages. First, the raw data are processed locally using the {\sc FULU} and VALC algorithms discussed in the following sections. Then, a dedicated script uploads the results through the service API, mapping them into internal data structures and storing them in the database. This approach enables rapid iteration over algorithm versions and allows updates without redeploying the service itself. However, as discussed below, this scheme becomes less suitable for our future development plans, and we outline the planned improvements in Section~\ref{sec:prosp}.

\begin{figure}
    \centering
    \includegraphics[width=\linewidth]{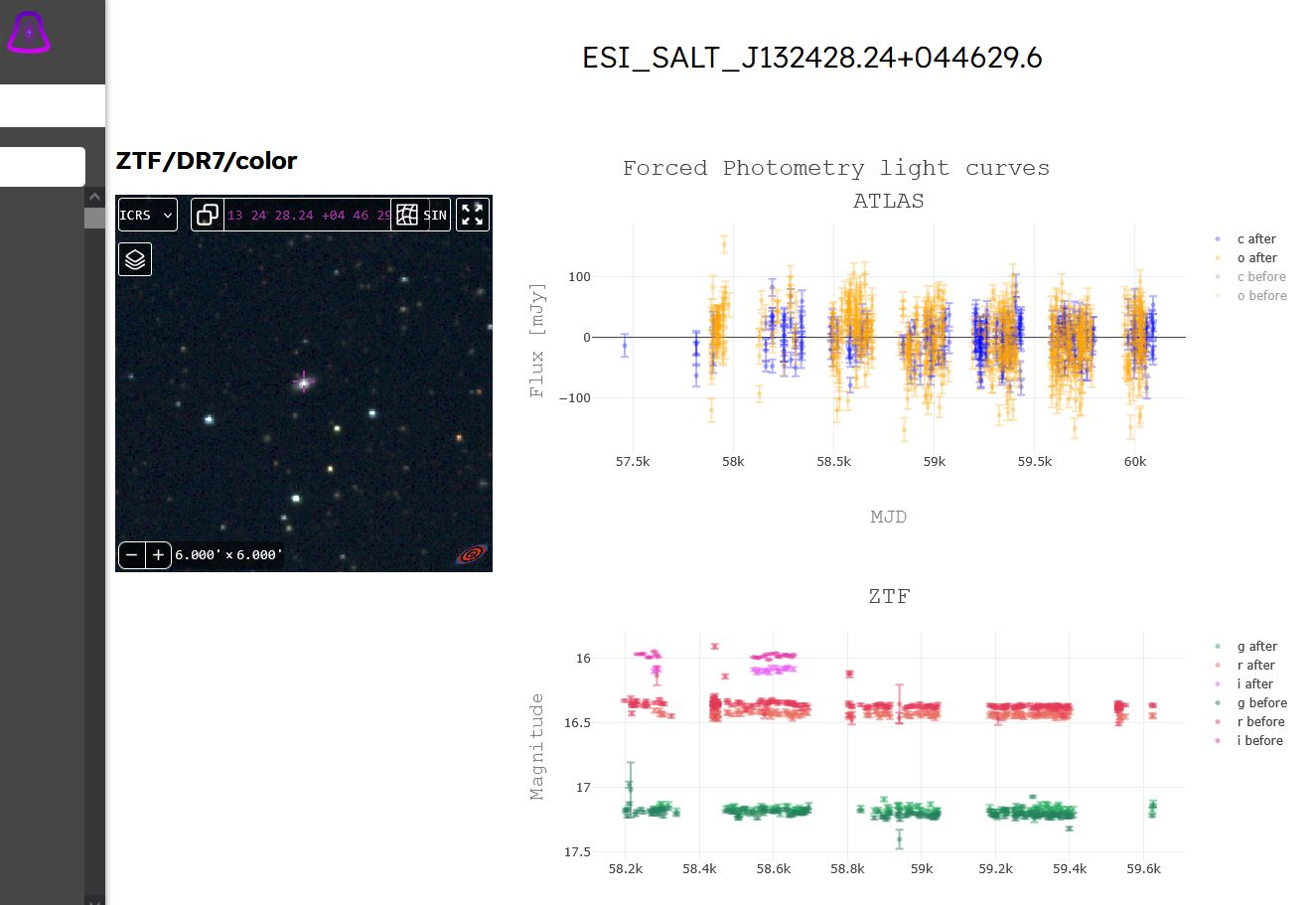}
    \caption{Interface example of the low-mass AGN light curves visualization at the VALC page.}
    \label{fig: screen_valc}
\end{figure}

The VALC page shows the list of objects from the $\sim$1900 low-mass AGN sample. The search is available, and in case of partial name input, it shows subsamples that satisfy this part of the IAU name of the object. The page of a particularly chosen object shows equatorial coordinates \texttt{ra} and \texttt{dec}; black hole mass measured using the broad H$\alpha$ component in SDSS (\texttt{MBH\_SDSS}), MagE (\texttt{MBH\_MagE}), ESI (\texttt{MBH\_ESI}), and SALT RSS (\texttt{MBH\_SALT}) spectra, with numeration if available several spectra from the same instrument~\citep{2024maeu.conf..155G,2022aems.conf..367G}; \texttt{source} of X-ray detection, photon index \texttt{Gamma}, hydrogen column density \texttt{NH} and X-ray luminosity \texttt{L\_X} from the X-ray power-law with photoelectric absorption fit, if any~\citep{2022aems.conf..304T}; \texttt{rcsed} - the link with coordinate search in the development version of the RCSEDv2\footnote{\url{https://dev-rcsed2.voxastro.org/}} catalog~\citep{2024ASPC..535..371R,2024ASPC..535..175G,2024ASPC..535..243K,2024ASPC..535..179C,2024ASPC..535..375K,2024ASPC..535..415T}. All other fields are introduced for internal use within the VOxAstro team.
The light curves from AFPS and ZFPS are shown both before/after post-processing. The post-processing for ZFPS light curves includes zero-point and color correction, filtering following ZTF guidelines, and subtraction of the median star from the image~\citep{2024ASPC..535..283D}. We show the {\sc Aladin-lite} interactive image viewer~\citep{2014ascl.soft02005B} of the ZTF catalog to indicate whether a target is within the ZTF sample. ZFPS light curves are visualized in $\{g,r,i\}$ passbands if any of the points were available after post-processing. The color correction term effectively converts ZTF magnitudes to the Pan-STARRS1~\citep[PS1;][]{2016arXiv161205560C} magnitudes. Therefore, the shift between `before' and `after' light curves is present. AFPS light curves underwent only median reference star subtraction and represent flux in $\{o,c\}$ passbands. Figure~\ref{fig: screen_valc} shows an example of the VALC webpage interface.

\begin{figure}
    \centering
    \includegraphics[width=0.85\linewidth]{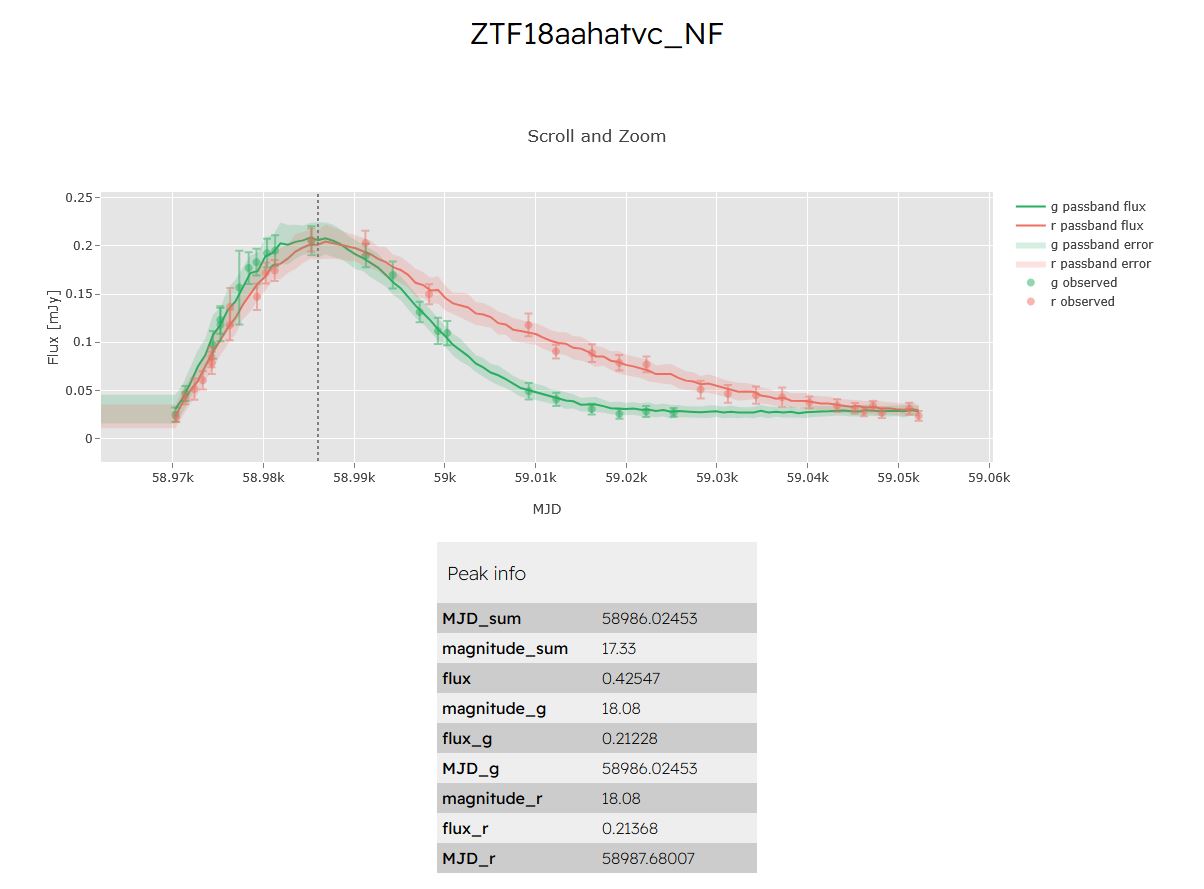}
    \caption{Interface example of the transient event light curve, approximated by normalizing flows at the {\sc Fulu} page.}
    \label{fig: screen_fulu}
\end{figure}

The {\sc Fulu} page shows a list of events from the BTS ZTF. The search by ZTF alert ID or part of this ID is also available. The page of the object represents the ZTF light curve in $\{g,r\}$ passbands, as well as peak parameters: (\texttt{MJD\_sum} - Modified Julian Date (MJD) of the peak, estimated from the sum of both filters light curve; \texttt{magnitude\_sum} - apparent magnitude of the peak calculated from the same cumulative light curve; \texttt{flux} - peak flux in $\mathrm{mJy}$ of the same cumulative light curve; \texttt{magnitude\_g}(\texttt{magnitude\_r}) - peak apparent magnitude in $g$($r$) passband light curve; \texttt{flux\_g}(\texttt{flux\_r}) - peak flux in $\mathrm{mJy}$; \texttt{MJD\_g}(\texttt{MJD\_r}) - MJD of the $g$($r$) peak in the assumption that we have one peak. Figure~\ref{fig: screen_fulu} shows an example of the {\sc Fulu} webpage interface.

The open-source code in the GitHub repository \footnote{\url{https://github.com/salamantos/FLEX}} facilitates the implementation of a similar approach in other projects and publications within the scientific community.

\section{Results}
The {\sc Fulu} page presents results of BTS ZTF light curves approximation by neural networks and GP~\citep{2023A&A...677A..16D,2023yCat..36770016D}. 
The VALC page of the website is used within the VOxAstro team and developed as a branch of the RCSEDv2 project. 
The dedicated follow-up programs were designed using the advantages of visualization within the VALC sample~\citep{2022hst..prop17177C,2022hst..prop17239C}. Particularly, monitoring programs resulted in the first photometric broad-line region (BLR) reverberation mapping campaign of a highly accreting intermediate-mass black hole~\citep[IMBH;][]{2025arXiv250503890D}.

\section{Prospects}
\label{sec:prosp}
We are preparing to release a web service that allows users to post-process ZTF and ATLAS light curves of AGNs. Optionally, the user could fit them using Brownian noise (non-stationary) or a damped random walk (DRW or Ornstein–Uhlenbeck process, requires stationarity), as implemented in the {\sc Fulu} Python package~\citep{2023A&A...677A..16D} by GP with a Matern1/2 kernel. Another option is to measure inter-band delays with linear interpolation of the cross-correlation function (CCF, \citet{1986ApJ...305..175G,1987ApJS...65....1G}), implemented in the {\sc pyCCF} Python package \citep{2018ascl.soft05032S}.
The plans also include publishing light curves in the German Astrophysical Virtual Observatory (GAVO) data centre~\citep{2007AN....328..713D}.

Such a service will require a different system architecture, as the raw data processing will need to occur on the server side. It will be necessary to define a common and convenient format for raw data uploads, which users will use to submit their objects. Additionally, a new backend module will be introduced to handle the execution of our algorithms on user-provided data. To prevent system overload, a message queue will likely be required to control and limit the incoming flow of user requests.

Regarding the frontend, as previously mentioned, the current implementation has become difficult to maintain. At present, it is written in plain HTML and JavaScript, and the project suffers from poor code structure. As a result, adding new logic or reusing existing components is complicated. We plan to migrate the frontend to {\sc React}~\citep{react}, which will allow modular component-based development and easier extension of functionality without affecting other parts of the system. Interaction with user input will also become more efficient, eliminating the need to manually attach event handlers to each element and process them elsewhere in the code. Overall, transitioning from legacy technologies to a modern framework will significantly improve maintainability and scalability.

\section*{\small Acknowledgements}
\scriptsize{MD expresses gratitude to Ivan Katkov, who shared VOxAstro {\sc Figma} templates to use as a basis for the VALC logo; the VOxAstro team, who shared the data for structurization and proper representation; Anton Afanasiev, who commented on the text clarity; Franz Bauer, who, with the ALeRCE team, introduced the use of filtering for ZFPS light curves following ZTF guidelines. Part of this work was done during the MD's internship at the Optical and Infrared Astronomy division of the Center for Astrophysics --- Harvard and Smithsonian. MD also acknowledges the Hubble Space Telescope grant HST-GO-16739 (PI: IC), which covered the internship in the Optical and Infrared (OIR) Division of the Smithsonian Astrophysical Observatory.
IC's research is supported by the Telescope Data Center at the Smithsonian Astrophysical Observatory. IC also acknowledges the support from the NASA ADAP-22-0102 grant (award 80NSSC23K0493) and NASA XMM-Newton Data Analysis grant (award 80NSSC22K0389). 
This work made use of Python packages SciPy \citep{2020SciPy-NMeth} and {\tt\string scikit-learn} \citep{scikit-learn}.
This research used Flexible Image Transport System (FITS) data format \citep{2001A&A...376..359H}. We complement the use of LLM-based tools (ChatGPT for finding appropriate wording and references in some cases, Grammarly for grammar and spelling convention checks) with critical thinking, as was proposed in \citep{2024arXiv240920252F}. The ZTF forced-photometry service was funded under the Heising-Simons Foundation grant \#12540303 (PI: Graham).}

\scriptsize
\bibliographystyle{ComBAO}
\nocite{*}
\bibliography{references}

@ARTICLE{2024arXiv240920252F,
       author = {{Fouesneau}, Morgan and {Momcheva}, Ivelina G. and {Chadayammuri}, Urmila and {Demianenko}, Mariia and {Dumont}, Antoine and {Hviding}, Raphael E. and {Kahle}, K. Angelique and {Pulatova}, Nadiia and {Rajpoot}, Bhavesh and {Scheuck}, Marten B. and {Seeburger}, Rhys and {Semenov}, Dmitry and {Villase{\~n}or}, Jaime I.},
        title = "{What is the Role of Large Language Models in the Evolution of Astronomy Research?}",
      journal = {arXiv e-prints},
         year = 2024,
        month = sep,
          eid = {arXiv:2409.20252},
        pages = {arXiv:2409.20252},
          doi = {10.48550/arXiv.2409.20252},
archivePrefix = {arXiv},
       eprint = {2409.20252},
 primaryClass = {astro-ph.IM},
       adsurl = {https://ui.adsabs.harvard.edu/abs/2024arXiv240920252F}
}

@ARTICLE{2001A&A...376..359H,
       author = {{Hanisch}, R.~J. and {Farris}, A. and {Greisen}, E.~W. and {Pence}, W.~D. and {Schlesinger}, B.~M. and {Teuben}, P.~J. and {Thompson}, R.~W. and {Warnock}, III, A.},
        title = "{Definition of the Flexible Image Transport System (FITS)}",
      journal = {\aap},
         year = 2001,
        month = sep,
       volume = {376},
        pages = {359-380},
          doi = {10.1051/0004-6361:20010923},
       adsurl = {https://ui.adsabs.harvard.edu/abs/2001A&A...376..359H}
}

@article{scikit-learn,
  title={Scikit-learn: Machine Learning in {P}ython},
  author={Pedregosa, F. and Varoquaux, G. and Gramfort, A. and Michel, V.
          and Thirion, B. and Grisel, O. and Blondel, M. and Prettenhofer, P.
          and Weiss, R. and Dubourg, V. and Vanderplas, J. and Passos, A. and
          Cournapeau, D. and Brucher, M. and Perrot, M. and Duchesnay, E.},
  journal={Journal of Machine Learning Research},
  volume={12},
  pages={2825--2830},
  year={2011}
}

@ARTICLE{2020SciPy-NMeth,
  author  = {Virtanen, Pauli and Gommers, Ralf and Oliphant, Travis E. and
            Haberland, Matt and Reddy, Tyler and Cournapeau, David and
            Burovski, Evgeni and Peterson, Pearu and Weckesser, Warren and
            Bright, Jonathan and {van der Walt}, St{\'e}fan J. and
            Brett, Matthew and Wilson, Joshua and Millman, K. Jarrod and
            Mayorov, Nikolay and Nelson, Andrew R. J. and Jones, Eric and
            Kern, Robert and Larson, Eric and Carey, C J and
            Polat, {\.I}lhan and Feng, Yu and Moore, Eric W. and
            {VanderPlas}, Jake and Laxalde, Denis and Perktold, Josef and
            Cimrman, Robert and Henriksen, Ian and Quintero, E. A. and
            Harris, Charles R. and Archibald, Anne M. and
            Ribeiro, Ant{\^o}nio H. and Pedregosa, Fabian and
            {van Mulbregt}, Paul and {SciPy 1.0 Contributors}},
  title   = {{{SciPy} 1.0: Fundamental Algorithms for Scientific
            Computing in Python}},
  journal = {Nature Methods},
  year    = {2020},
  volume  = {17},
  pages   = {261--272},
  adsurl  = {https://rdcu.be/b08Wh},
  doi     = {10.1038/s41592-019-0686-2},
}

@ARTICLE{2025arXiv250503890D,
       author = {{Demianenko}, Mariia and {Afanasiev}, Anton and {Rubtsov}, Evgenii and {Toptun}, Victoria and {Pott}, J{\"o}rg-Uwe and {Belinski}, Alexandr and {Bauer}, Franz and {Chilingarian}, Igor and {Grishin}, Kirill and {Burlak}, Marina and {Ikonnikova}, Natalia},
        title = "{Broad line region echo from highly accreting intermediate-mass black hole candidate SDSS J144850.08+160803.1. The first probe of intra-night variability and reverberation mapping}",
      journal = {arXiv e-prints},
         year = 2025,
        month = may,
          eid = {arXiv:2505.03890},
        pages = {arXiv:2505.03890},
          doi = {10.48550/arXiv.2505.03890},
archivePrefix = {arXiv},
       eprint = {2505.03890},
 primaryClass = {astro-ph.GA},
       adsurl = {https://ui.adsabs.harvard.edu/abs/2025arXiv250503890D}
}

@article{merkel2014docker,
  title={Docker: lightweight linux containers for consistent development and deployment},
  author={Merkel, Dirk},
  journal={Linux journal},
  volume={2014},
  number={239},
  pages={2},
  year={2014}
}

@misc{fastapi,
  author = {Sebastián Ramírez},
  title = {FastAPI},
  year = {2018},
  url = {https://fastapi.tiangolo.com/}
}

@misc{mongodb,
  author = {MongoDB, Inc.},
  title = {MongoDB},
  year = {2009},
  url = {https://www.mongodb.com}
}

@manual{react,
  title        = {{React}: A JavaScript library for building user interfaces},
  author       = {{Meta Platforms, Inc. \& contributors}},
  year         = {2025},
  url          = {https://react.dev/},
  note         = {Version 18 or later}
}

@misc{crockford2006json,
  author       = {Douglas Crockford},
  title        = {The application/json Media Type for JavaScript Object Notation (JSON)},
  year         = {2006},
  howpublished = {\url{https://www.ietf.org/rfc/rfc4627.txt}}
}

@online{plotly, 
author = {Plotly Technologies Inc.},
title = {Collaborative data science},
publisher = {Plotly Technologies Inc.},
address = {Montreal, QC},
year = {2015},
url = {https://plot.ly} }

@INPROCEEDINGS{2024ASPC..535..283D,
       author = {{Demianenko}, M. and {Grishin}, K. and {Toptun}, V. and {Chilingarian}, I. and {Katkov}, I. and {Goradzhanov}, V. and {Kuzmin}, I.},
        title = "{Optical Variability of ``Light-weight'' Supermassive Black Holes at a Few Percent Level from ZTF Forced-Photometry Light Curves}",
    booktitle = {Astromical Data Analysis Software and Systems XXXI},
         year = 2024,
       editor = {{Hugo}, B.~V. and {Van Rooyen}, R. and {Smirnov}, O.~M.},
       series = {Astronomical Society of the Pacific Conference Series},
       volume = {535},
        month = may,
        pages = {283},
          doi = {10.48550/arXiv.2201.03712},
archivePrefix = {arXiv},
       eprint = {2201.03712},
 primaryClass = {astro-ph.GA},
       adsurl = {https://ui.adsabs.harvard.edu/abs/2024ASPC..535..283D}
}

@ARTICLE{2023A&A...677A..16D,
       author = {{Demianenko}, Mariia and {Malanchev}, Konstantin and {Samorodova}, Ekaterina and {Sysak}, Mikhail and {Shiriaev}, Aleksandr and {Derkach}, Denis and {Hushchyn}, Mikhail},
        title = "{Understanding of the properties of neural network approaches for transient light curve approximations}",
      journal = {\aap},
         year = 2023,
        month = sep,
       volume = {677},
          eid = {A16},
        pages = {A16},
          doi = {10.1051/0004-6361/202245189},
archivePrefix = {arXiv},
       eprint = {2209.07542},
 primaryClass = {astro-ph.IM},
       adsurl = {https://ui.adsabs.harvard.edu/abs/2023A&A...677A..16D}
}

@ARTICLE{2019PASP..131a8003M,
       author = {{Masci}, Frank J. and {Laher}, Russ R. and {Rusholme}, Ben and {Shupe}, David L. and {Groom}, Steven and {Surace}, Jason and {Jackson}, Edward and {Monkewitz}, Serge and {Beck}, Ron and {Flynn}, David and {Terek}, Scott and {Landry}, Walter and {Hacopians}, Eugean and {Desai}, Vandana and {Howell}, Justin and {Brooke}, Tim and {Imel}, David and {Wachter}, Stefanie and {Ye}, Quan-Zhi and {Lin}, Hsing-Wen and {Cenko}, S. Bradley and {Cunningham}, Virginia and {Rebbapragada}, Umaa and {Bue}, Brian and {Miller}, Adam A. and {Mahabal}, Ashish and {Bellm}, Eric C. and {Patterson}, Maria T. and {Juri{\'c}}, Mario and {Golkhou}, V. Zach and {Ofek}, Eran O. and {Walters}, Richard and {Graham}, Matthew and {Kasliwal}, Mansi M. and {Dekany}, Richard G. and {Kupfer}, Thomas and {Burdge}, Kevin and {Cannella}, Christopher B. and {Barlow}, Tom and {Van Sistine}, Angela and {Giomi}, Matteo and {Fremling}, Christoffer and {Blagorodnova}, Nadejda and {Levitan}, David and {Riddle}, Reed and {Smith}, Roger M. and {Helou}, George and {Prince}, Thomas A. and {Kulkarni}, Shrinivas R.},
        title = "{The Zwicky Transient Facility: Data Processing, Products, and Archive}",
      journal = {\pasp},
         year = 2019,
        month = jan,
       volume = {131},
       number = {995},
        pages = {018003},
          doi = {10.1088/1538-3873/aae8ac},
archivePrefix = {arXiv},
       eprint = {1902.01872},
 primaryClass = {astro-ph.IM},
       adsurl = {https://ui.adsabs.harvard.edu/abs/2019PASP..131a8003M}
}

@ARTICLE{2023arXiv230516279M,
       author = {{Masci}, Frank J. and {Laher}, Russ R. and {Rusholme}, Benjamin and {Shupe}, David and {Paladini}, Roberta and {Groom}, Steve and {Wold}, Avery and {Miller}, Adam A. and {Drake}, Andrew},
        title = "{A New Forced Photometry Service for the Zwicky Transient Facility}",
      journal = {arXiv e-prints},
         year = 2023,
        month = may,
          eid = {arXiv:2305.16279},
        pages = {arXiv:2305.16279},
          doi = {10.48550/arXiv.2305.16279},
archivePrefix = {arXiv},
       eprint = {2305.16279},
 primaryClass = {astro-ph.IM},
       adsurl = {https://ui.adsabs.harvard.edu/abs/2023arXiv230516279M}
}

@INPROCEEDINGS{2022aems.conf..359D,
       author = {{Demianenko}, M. and {Grishin}, K. and {Toptun}, V. and {Chilingarian}, I. and {Katkov}, I. and {Goradzhanov}, V. and {Kuzmin}, I.},
        title = "{Optical light curves of light-weight supermassive black holes produced by the Zwicky Transient Facility Forced Photometry Service}",
    booktitle = {Astronomy at the Epoch of Multimessenger Studies},
         year = 2022,
        month = jan,
        pages = {359-361},
          doi = {10.51194/VAK2021.2022.1.1.144},
archivePrefix = {arXiv},
       eprint = {2112.11520},
 primaryClass = {astro-ph.GA},
       adsurl = {https://ui.adsabs.harvard.edu/abs/2022aems.conf..359D}
}

@INPROCEEDINGS{2024maeu.conf..155G,
       author = {{Goradzhanov}, V. and {Chilingarian}, I. and {Demianenko}, M. and {Katkov}, I. and {Grishin}, K. and {Toptun}, V. and {Rubtsov}, E. and {Gasymov}, D. and {Kuzmin}, I.},
        title = "{Optical spectroscopy of host-galaxies of intermediate mass black holes: evolution of central black holes}",
    booktitle = {Modern Astronomy: From the Early Universe to Exoplanets and Black Holes},
         year = 2024,
        month = dec,
        pages = {155-161},
          doi = {10.26119/VAK2024.022},
       adsurl = {https://ui.adsabs.harvard.edu/abs/2024maeu.conf..155G}
}

@INPROCEEDINGS{2022aems.conf..367G,
       author = {{Goradzhanov}, V. and {Chilingarian}, I. and {Katkov}, I. and {Grishin}, K. and {Toptun}, V. and {Kuzmin}, I. and {Demianenko}, M.},
        title = "{Optical spectroscopic observations of intermediate-mass black holes and their host galaxies: the MBH {\ensuremath{-}} {\ensuremath{\sigma}}{\ensuremath{*}} relation}",
    booktitle = {Astronomy at the Epoch of Multimessenger Studies},
         year = 2022,
        month = jan,
        pages = {367-369},
          doi = {10.51194/VAK2021.2022.1.1.147},
archivePrefix = {arXiv},
       eprint = {2201.01078},
 primaryClass = {astro-ph.GA},
       adsurl = {https://ui.adsabs.harvard.edu/abs/2022aems.conf..367G}
}

@ARTICLE{2018ApJ...863....1C,
       author = {{Chilingarian}, Igor V. and {Katkov}, Ivan Yu. and {Zolotukhin}, Ivan Yu. and {Grishin}, Kirill A. and {Beletsky}, Yuri and {Boutsia}, Konstantina and {Osip}, David J.},
        title = "{A Population of Bona Fide Intermediate-mass Black Holes Identified as Low-luminosity Active Galactic Nuclei}",
      journal = {\apj},
         year = 2018,
        month = aug,
       volume = {863},
       number = {1},
          eid = {1},
        pages = {1},
          doi = {10.3847/1538-4357/aad184},
archivePrefix = {arXiv},
       eprint = {1805.01467},
 primaryClass = {astro-ph.GA},
       adsurl = {https://ui.adsabs.harvard.edu/abs/2018ApJ...863....1C}
}

@ARTICLE{1986ApJ...305..175G,
       author = {{Gaskell}, C.~M. and {Sparke}, L.~S.},
        title = "{Line Variations in Quasars and Seyfert Galaxies}",
      journal = {\apj},
         year = 1986,
        month = jun,
       volume = {305},
        pages = {175},
          doi = {10.1086/164238},
       adsurl = {https://ui.adsabs.harvard.edu/abs/1986ApJ...305..175G}
}

@ARTICLE{1987ApJS...65....1G,
       author = {{Gaskell}, C. Martin and {Peterson}, Bradley M.},
        title = "{The Accuracy of Cross-Correlation Estimates of Quasar Emission-Line Region Sizes}",
      journal = {\apjs},
         year = 1987,
        month = sep,
       volume = {65},
        pages = {1},
          doi = {10.1086/191216},
       adsurl = {https://ui.adsabs.harvard.edu/abs/1987ApJS...65....1G}
}

@software{2018ascl.soft05032S,
       author = {{Sun}, Mouyuan and {Grier}, C.~J. and {Peterson}, B.~M.},
        title = "{PyCCF: Python Cross Correlation Function for reverberation mapping studies}",
 howpublished = {Astrophysics Source Code Library, record ascl:1805.032},
         year = 2018,
        month = may,
          eid = {ascl:1805.032},
       adsurl = {https://ui.adsabs.harvard.edu/abs/2018ascl.soft05032S}
}

@ARTICLE{2021TNSAN...7....1S,
       author = {{Shingles}, L. and {Smith}, K.~W. and {Young}, D.~R. and {Smartt}, S.~J. and {Tonry}, J. and {Denneau}, L. and {Heinze}, A. and {Weiland}, H. and {Flewelling}, H. and {Stalder}, B. and {Clocchiatti}, A. and {F{\"o}rster}, F. and {Pignata}, G. and {Rest}, A. and {Anderson}, J. and {Stubbs}, C. and {Erasmus}, N.},
        title = "{Release of the ATLAS Forced Photometry server for public use}",
      journal = {Transient Name Server AstroNote},
         year = 2021,
        month = jan,
       volume = {7},
        pages = {1-7},
       adsurl = {https://ui.adsabs.harvard.edu/abs/2021TNSAN...7....1S}
}

@INPROCEEDINGS{2022aems.conf..304T,
       author = {{Toptun}, V. and {Chilingarian}, I. and {Grishin}, K. and {Katkov}, I. and {Zolotukhin}, I. and {Goradzhanov}, V. and {Demianenko}, M. and {Kuzmun}, I.},
        title = "{Confirmation of intermediate-mass black holes candidates with x-ray observations}",
    booktitle = {Astronomy at the Epoch of Multimessenger Studies},
         year = 2022,
        month = jan,
        pages = {304-306},
          doi = {10.51194/VAK2021.2022.1.1.117},
archivePrefix = {arXiv},
       eprint = {2201.01075},
 primaryClass = {astro-ph.HE},
       adsurl = {https://ui.adsabs.harvard.edu/abs/2022aems.conf..304T}
}

@ARTICLE{2007AN....328..713D,
       author = {{Demleitner}, Markus and {Gufler}, Benjamin and {Kim}, Jaiwon and {Lemson}, Gerard and {Nickelt-Czycykowski}, Iliya and {Rauch}, Thomas and {Stampa}, Ulrike and {Steinmetz}, Matthias and {Voges}, Wolfgang and {Wambsganss}, Joachim},
        title = "{The German Astrophysical Virtual Observatory (GAVO): Archives and Applications, Status and Services}",
      journal = {Astronomische Nachrichten},
         year = 2007,
        month = sep,
       volume = {328},
       number = {7},
        pages = {713},
       adsurl = {https://ui.adsabs.harvard.edu/abs/2007AN....328..713D}
}

@INPROCEEDINGS{2024ASPC..535..371R,
       author = {{Rubtsov}, E. and {Chilingarian}, I. and {Katkov}, I. and {Grishin}, K. and {Goradzhanov}, V. and {Borisov}, S.},
        title = "{Hybrid minimization algorithm for computationally expensive multi-dimensional fitting}",
    booktitle = {Astromical Data Analysis Software and Systems XXXI},
         year = 2024,
       editor = {{Hugo}, B.~V. and {Van Rooyen}, R. and {Smirnov}, O.~M.},
       series = {Astronomical Society of the Pacific Conference Series},
       volume = {535},
        month = may,
        pages = {371},
          doi = {10.48550/arXiv.2112.03413},
archivePrefix = {arXiv},
       eprint = {2112.03413},
 primaryClass = {astro-ph.IM},
       adsurl = {https://ui.adsabs.harvard.edu/abs/2024ASPC..535..371R}
}

@INPROCEEDINGS{2024ASPC..535..175G,
       author = {{Goradzhanov}, V. and {Chilingarian}, I. and {Rubtsov}, E. and {Katkov}, I. and {Grishin}, K. and {Toptun}, V. and {Kasparova}, A. and {Klochkov}, V. and {Borisov}, S.},
        title = "{RCSEDv2: Processing and analysis of 4+ million galaxy spectra}",
    booktitle = {Astromical Data Analysis Software and Systems XXXI},
         year = 2024,
       editor = {{Hugo}, B.~V. and {Van Rooyen}, R. and {Smirnov}, O.~M.},
       series = {Astronomical Society of the Pacific Conference Series},
       volume = {535},
        month = may,
        pages = {175},
          doi = {10.48550/arXiv.2112.04865},
archivePrefix = {arXiv},
       eprint = {2112.04865},
 primaryClass = {astro-ph.IM},
       adsurl = {https://ui.adsabs.harvard.edu/abs/2024ASPC..535..175G}
}

@INPROCEEDINGS{2024ASPC..535..243K,
       author = {{Klochkov}, V. and {Katkov}, I. and {Chilingarian}, I. and {Grishin}, K. and {Kasparova}, A. and {Goradzhanov}, V. and {Toptun}, V. and {Rubtsov}, E. and {Borisov}, S.},
        title = "{RCSEDv2: Open-source web tools for visualization of imaging and spectral data}",
    booktitle = {Astromical Data Analysis Software and Systems XXXI},
         year = 2024,
       editor = {{Hugo}, B.~V. and {Van Rooyen}, R. and {Smirnov}, O.~M.},
       series = {Astronomical Society of the Pacific Conference Series},
       volume = {535},
        month = may,
        pages = {243},
          doi = {10.48550/arXiv.2112.04867},
archivePrefix = {arXiv},
       eprint = {2112.04867},
 primaryClass = {astro-ph.IM},
       adsurl = {https://ui.adsabs.harvard.edu/abs/2024ASPC..535..243K}
}

@software{2014ascl.soft02005B,
       author = {{Boch}, Thomas},
        title = "{Aladin Lite: Lightweight sky atlas for browsers}",
 howpublished = {Astrophysics Source Code Library, record ascl:1402.005},
         year = 2014,
        month = feb,
          eid = {ascl:1402.005},
archivePrefix = {ascl},
       eprint = {1402.005},
       adsurl = {https://ui.adsabs.harvard.edu/abs/2014ascl.soft02005B}
}

\newpage
\appendix
\renewcommand{\thesection}{\Alph{section}.\arabic{section}}
\setcounter{section}{0}
\normalsize

\end{document}